\documentclass[runningheads]{svmult}

\usepackage{amssymb}
\usepackage{makeidx}   
\usepackage{graphicx}  
\usepackage{subeqnar}  
\usepackage{multicol}  
\usepackage{physprbb}  
\makeindex             

\begin{document}

\title*{Metallic Ferromagnetism - an Electronic Correlation Phenomenon}
\toctitle{Metallic Ferromagnetism -- an Electronic Correlation
Phenomenon}
\titlerunning{Metallic Ferromagnetism -- an Electronic Correlation
Phenomenon}

\author{D.~Vollhardt\inst{1}
\and N.~Bl{\"u}mer\inst{1,*} \and K.~Held\inst{2} \and
M.~Kollar\inst{3}}
\authorrunning{D.~Vollhardt et al.}



\institute{Theoretical Physics III, Center for Electronic
Correlations and Magnetism, University of Augsburg, 86135
Augsburg, Germany \and Dept. of Physics, Princeton University,
Princeton, NJ 08544, USA \and Dept. of Physics, Yale
    University, P.~O.~Box 208120, New Haven, CT 06520-8120, USA
} \maketitle

\begin{abstract}
New insights into the microscopic origin of itinerant
ferromagnetism were recently gained from investigations of
electronic lattice models within dynamical mean-field theory
(DMFT). In particular, it is now established that even in the
one-band Hubbard model metallic ferromagnetism is stable at
intermediate values of the interaction $U$ and density $n$ on
regular, frustrated lattices. Furthermore, band degeneracy along
with Hund's rule couplings is very effective in stabilizing
metallic ferromagnetism in a broad range of electron fillings.
DMFT also permits one to investigate more complicated correlation
models, e.g., the ferromagnetic Kondo lattice model with Coulomb
interaction, describing electrons in manganites with perovskite
structure. Here we review recent results obtained with DMFT which
help to clarify the origin of band-ferromagnetism as a correlation
phenomenon.
\end{abstract}

\section{Introduction}

Since the Curie temperature of ferromagnetic metals like iron,
cobalt, and nickel is of the order of electrostatic energies in
solids, i.e., is much higher than the interaction energy of the
electron spins, itinerant ferromagnetism is expected to be the
result of the interplay between the ordinary spin-{\it
in}dependent Coulomb interaction and
the kinetic energy of the electrons, in combination with the Pauli principle %
\cite{Ashcroft,Mattis88,Fazekas}. As such it is one of the
fundamental cooperative phenomena in condensed matter physics.

Until recently the theory of itinerant ferromagnetism was
investigated by two essentially separate communities, one using
model Hamiltonians in conjunction with many-body techniques (or
even rigorous mathematical methods
\cite{Lieb62,LiebFaz,Mielke-Tasaki}), the other employing density
functional theory (DFT) \cite{Hohenberg64,Moruzzi}. DFT
and its local density approximation (LDA) have the advantage of being {\em %
ab initio} approaches which do not require empirical parameters as
input. Indeed, they are highly successful techniques for the
calculation of the electronic structure of real materials.
However, in practice DFT/LDA is restricted in its ability to
describe strongly correlated materials. Here, the model
Hamiltonian approach is more general and powerful since there
exist systematic theoretical methods to investigate the
many-electron problem with increasing accuracy. Nevertheless, the
uncertainty in the choice of the model parameters and the
technical complexity of the correlation problem itself prevent the
model Hamiltonian approach from being a flexible or reliable
enough tool for studying real materials. The two approaches are
therefore complementary.

Originally the one-band Hubbard model was introduced to gain
insight into the origin of metallic ferromagnetism
\cite{Gutzwiller63,Hubbard63,Kanamori63}. However, even for this
simplest possible microscopic model answers are not easily
obtained
since in general metallic ferromagnetism occurs at {\it intermediate }%
couplings and off half filling \cite{Fazekas,Vollhardt97b+99}.
Thus, this cooperative phenomenon belongs to the class of problems
where standard perturbation techniques are not applicable. In
particular, weak-coupling theories or renormalization group
approaches \cite{Shankar94} which are so effective in detecting
instabilities with respect to antiferromagnetism or
superconductivity meet with limited success here. In general, {\it
non-perturbative} investigations are required.

During the last couple of years significant progress was made in
the theoretical understanding of the microscopic foundations of
metallic ferromagnetism. They were made possible mainly by the
development (i)\ of new analytic approaches, such as the
mathematical methods used to investigate flat-band ferromagnetism
\cite{Mielke-Tasaki} and its extensions \cite{Penc96,Pieri96} as
well as dynamical mean-field theory (DMFT)
\cite{Metzner89a,Uhrig96,Hanisch97}, (ii) of new numerical
techniques, such as the density matrix renormalization group
(DMRG) which yields precise results in $d=1$ \cite{Daul}, and
(iii) of new comprehensive approximation schemes such as the
multi-band Gutzwiller wave function \cite{Weber}, or the new {\em
ab initio} computational scheme LDA+DMFT
\cite{Anisimov,Lichtenstein,Held00} which merges conventional band
structure theory (LDA) with a comprehensive many-body technique
(DMFT).

In this paper the insights gained with the help of the DMFT will
be reviewed. After a short introduction into this approach
(Section \ref{DMFT}), the microscopic conditions for metallic
ferromagnetism in the one-band Hubbard model (Section
\ref{sec:oneband}) and in the case of the orbitally degenerate
model (Section \ref{sec:zweiband}) are explained and the
differences discussed. Furthermore, the physics of itinerant
ferromagnetism in more complicated models, e.g., the ferromagnetic
Kondo lattice model with Coulomb correlations for
manganites with perovskite structure, is analyzed. A conclusion
and outlook (Section \ref{sec:concl}) closes the presentation.


\section{Dynamical Mean-Field Theory}

\label{DMFT}

During the last decade a new many-body approach was developed
which is especially well-suited for the investigation of
correlated electronic systems with strong local interactions --
the dynamical mean-field theory (DMFT)~\cite{DMFT}. It becomes
exact for $d=\infty $, i.e., for lattices with coordination number
$Z=\infty$ \cite{Metzner89a}. Why is the investigation of this
limit useful and interesting? The answer is that already in $d=3$
the coordination number of regular lattices, such as the fcc
lattice, is quite large ($Z_{\mbox{\scriptsize fcc}}=12$).  It is
therefore quite natural to view $Z$ as a large number, and to
consider the limit
$Z\rightarrow \infty$ as a starting point \cite{Metzner89a}. To
obtain a meaningful model in this limit one has to scale the NN~
hopping amplitude in the kinetic energy (see below) as $t=t^{\ast
}/\sqrt{Z}$ (in the following we set $t^{\ast }=1$ and thereby fix
the energy scale). Then one obtains a purely local theory where
the self-energy $\Sigma _{{\bf k}}(\omega )$ becomes ${\bf k}$
independent and where the propagator $G({\bf k},\omega
)=G^{0}({\bf k},\omega -\Sigma (\omega ))$ may be represented by
the non-interacting propagator at a shifted frequency~\cite{DMFT}.
The dynamics of the quantum mechanical correlation problem is then
fully included. That is why this theory is referred to as
``dynamical mean-field theory'' (DMFT).  We note that due to the
local nature of the theory there is no {\it short}-range order in
position space.

Within DMFT the electronic lattice problem is mapped onto an
effective self-consistent single-impurity Anderson model
\cite{DMFT}. DMFT is a non-perturbative, thermodynamically
consistent theoretical framework within which the dynamics of
correlated lattice electrons with local interaction can be
investigated at all coupling strengths. This is of essential
importance for problems like band-ferromagnetism or the
metal-insulator transition \cite{MIT} which set in at a Coulomb
interaction strength comparable to the electronic band width.

In DMFT the information about the lattice or the dispersion of the
system under investigation enters only through the density of
states (DOS) $N^{0}(E)$ of the {\it non}-interacting particles,
unless there is long-range order with broken translational
symmetry of the lattice as in the case of antiferromagnetism. In
finite dimensions, e.g., $d=3$, DMFT is then an approximation --
usually an excellent one as is manifested by the plethora of
results obtained within the last decade \cite{DMFT}. DMFT is
presently the only comprehensive, thermodynamically consistent
computational scheme which allows one to investigate the dynamics
of correlated lattice electrons on all energy scales.

Due to its equivalence to an Anderson impurity problem a variety
of approximative techniques have been employed to solve the
complicated DMFT equations, such as the iterated perturbation
theory (IPT) \cite{georges92} and the non-crossing approximation
(NCA) \cite{NCA2}, as well as numerically exact techniques like
quantum Monte-Carlo simulations (QMC) \cite{QMC}, exact
diagonalization (ED) \cite{caffarel94}, or the numerical
renormalization group (NRG) \cite{NRG}. However, NRG cannot yet be
used to solve the DMFT equations for multi-band models.

In the present paper all DMFT results were obtained by QMC. This
is a particularly well-tested and reliable, albeit
computer-expensive method which may be employed down to
temperatures $T\sim 10^{-2}W$ ($W$: band width) and at not too
strong interactions.

\section{The One-Band Hubbard Model}

\label{sec:oneband}The one-band Hubbard model%
\begin{equation}  \label{eqn:hub}
{H}_{\mbox{\scriptsize Hub}}=-\sum_{i,j,\sigma
}t_{ij}({c}_{i\sigma
}^{\dagger }{c}_{j\sigma }^{\phantom\dagger }+{\rm h.c.})+U\sum_{i}{n}%
_{i\uparrow }{n}_{i\downarrow }
\end{equation}%
is the simplest lattice model for correlated electrons, and was
originally proposed to understand metallic ferromagnetism in 3$d$
transition metals \cite{Gutzwiller63,Hubbard63,Kanamori63}. Here,
$t_{ij}$ is a general hopping matrix element between sites $i$ and
$j$. In the past the kinetic energy in (\ref{eqn:hub}) was usually
restricted to nearest-neighbor (NN) hopping; then it is useful to
divide the underlying lattices into bipartite and non-bipartite
ones. This distinction ceases to be useful if, in addition to NN~
hopping $t$, longer-range hopping, in particular
next-nearest-neighbor (NNN) hopping $t^{\prime }$
is considered.

The Hubbard model is characterized by a purely local interaction
term which is completely independent of the lattice structure and
dimension. Therefore in this model the kinetic energy of the
electrons (i.e., the dispersion) and the structure of the lattice
are very important for the stability of metallic ferromagnetism.
This is well-known from the approximate investigations by
Gutzwiller \cite{Gutzwiller63}, Hubbard \cite{Hubbard63}, Kanamori
\cite{Kanamori63} and Nagaoka \cite{Nagaoka66}, and has recently
been confirmed, and made precise, by detailed investigations
\cite{Mielke-Tasaki,MuellerHartmann95,Pieri96,Penc96,Uhrig96,Hanisch97}.

\subsection{Routes to Ferromagnetism}
\label{subs:routes}

On bipartite lattices the $t^{\prime }$-hopping term destroys the
antiferromagnetic nesting instability at small $U$ (see, for example \cite%
{Hofstetter98}) by shifting spectral weight to the band edges and
thereby introducing an asymmetry into the otherwise symmetric DOS.
It will be shown
below that a high spectral weight at the band edge (more precisely: the {\it %
lower} band edge for $n<1$) minimizes the loss of kinetic energy
of the overturned spins in the magnetic state and is hence
energetically favorable. Therefore frustrated, i.e.,~non-bipartite
lattices, or bipartite lattices with frustration due to
further-range hopping (e.g.,~$t^{\prime }\not=0$) support the
stabilization of metallic ferromagnetism.

The fcc lattice is an example of a frustrated lattice in $d=3$.
The
corresponding DOS of the non-interacting particles is shown in Fig.~\ref%
{fig:fccdos}. Switching on an additional NNN~hopping $t^{\prime }$
further increases the spectral weight at the lower band edge. For $%
t^{\prime }=t/2$ one even obtains a square-root-like divergence.
\begin{figure}[tbp]
\begin{center}
\includegraphics[width=0.6\hsize]{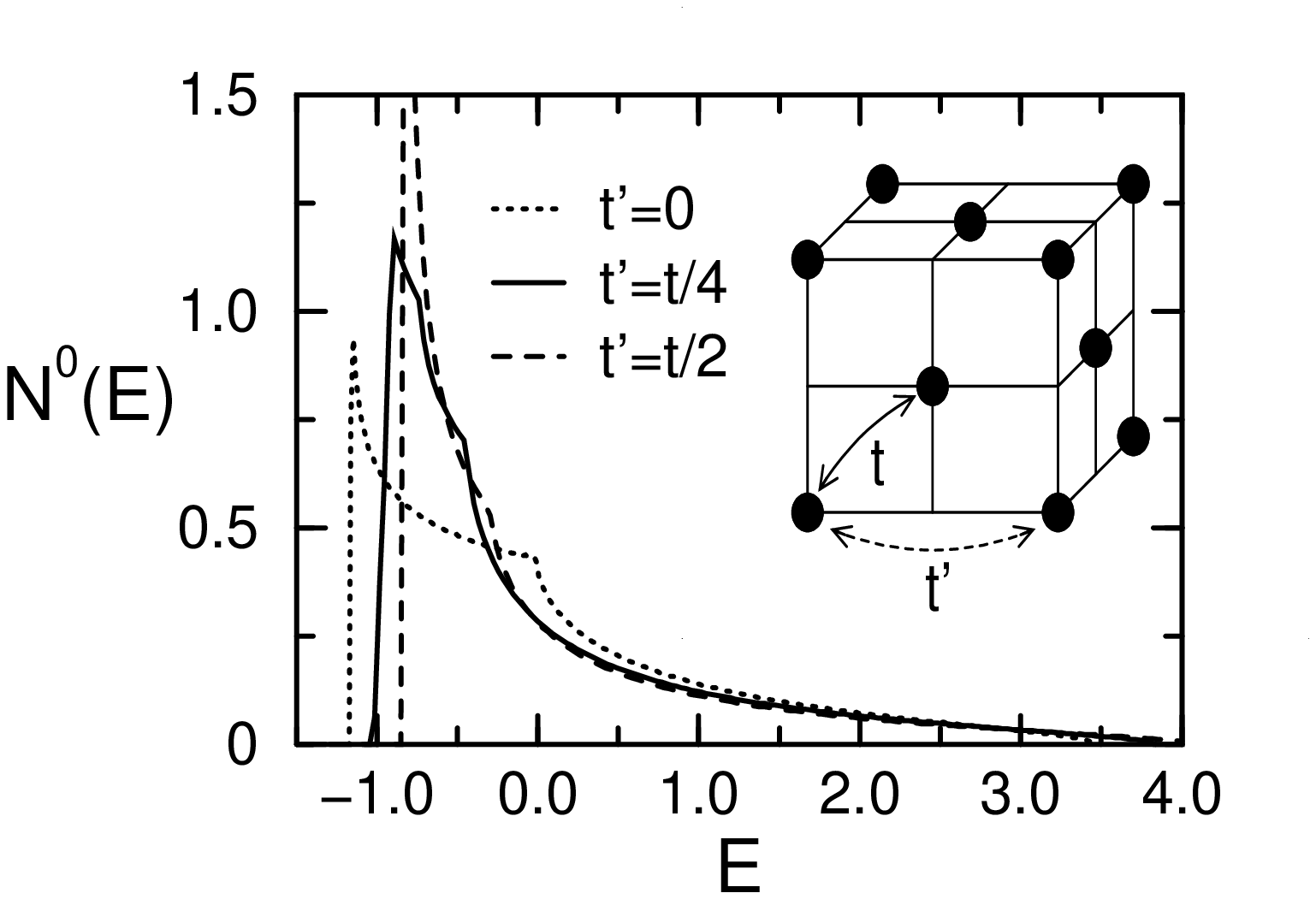}
\end{center}
\caption{DOS of noninteracting electrons on a fcc lattice in $d=3$
with and without additional NNN~hopping $t^{\prime}$.}
\label{fig:fccdos}
\end{figure}
To understand why a high spectral weight at the band edge is
favorable for
the kinetic energy we first consider the case $U=0$, $n<1$~\cite%
{Vollhardt97b+99}. Let us consider a flat, symmetric DOS. The
fully polarized state is obtained by inverting the spin of the
down electrons, which due to the Pauli principle have to occupy
higher energy states. Counting the energy from the lower band edge
the Fermi energy of the polarized state, $\mu _{\uparrow }$, is
seen to be twice that of the unpolarized state. This should be
contrasted with a DOS having large spectral weight at the lower
band edge. Here the Fermi level of the polarized state is not so
strongly shifted upwards, i.e.,~fewer high energy states are
populated, which is clearly energetically favorable. The energy
difference between the fully polarized state and the unpolarized
state
\begin{equation}
\Delta E=\bigg[\int\limits_{-W_{1}}^{\mu _{\uparrow
}}-\;2\int\limits_{-W_{1}}^{\mu }\bigg]\;dE\;N^{0}(E)\;E
\end{equation}%
must become negative for the ferromagnetic state to be stable. Of
course, in the non-interacting case one has $\Delta
E>0$~\cite{Lieb62}. Nevertheless, even for $U=0$, $\Delta E$
attains its lowest value for a DOS with peaked spectral weight at
the lower band edge for all
$n<1$~\cite{Vollhardt97b+99}. To show that $\Delta E<0$ for $U>0$
requires a reliable calculation of the energy of the {\it correlated
  paramagnet}. Indeed, this is one of the central, most difficult
problems of electronic correlation theory. It should be noted that the
above discussion concerning the shape of the DOS goes beyond the
well-known Stoner criterion which predicts a ferromagnetic instability
of the paramagnet for $U$ equal to the inverse of the DOS precisely
{\it at} the Fermi level.


\subsection{Frustrated Lattices}

\label{sec:asymmetricDOS}

Since metallic ferromagnetism is an intermediate-coupling problem
purely analytic approaches alone cannot provide sufficient
information, in particular in dimensions $d>1$. In this situation
the development of new numerical techniques during the last few
years was of crucial importance for progress in this field. In
particular, for one-dimensional systems the DMRG \cite{Daul}, and
for high-dimensional systems the DMFT have led to important
insights. Here we restrict ourselves to results obtained by DMFT.

By a suitable generalization of the dispersion relation in three
dimensions, frustrated lattices like the fcc lattice can be
defined in any dimension, in particular in $d=\infty $
\cite{MH91}. With the proper scaling of the hopping term (see
above) the non-interacting DOS of the {\it g}eneralized {\it fcc}
lattice in $d=\infty $ is obtained as:
\begin{equation}
N_{\mbox{\scriptsize\em gfcc}}^{0}(E)=e^{-(1+\sqrt{2}E)/2}/\sqrt{\pi (1+%
\sqrt{2}E)}  \label{glferro1}
\end{equation}%
It has a strong square-root divergence at the lower band edge, $-1/\sqrt{2}$%
, and no upper band edge.

\begin{figure}[tbp]
\begin{center}
\unitlength1cm
\begin{picture}(6.5,4.4)
\put(0,-0.3){\includegraphics[width=0.5\hsize]{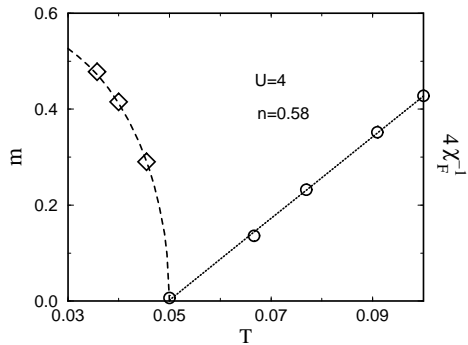}}
\end{picture}
\end{center}
\caption{ Magnetization $m$ (diamonds) and inverse ferromagnetic
susceptibility $\protect\chi _{F}^{-1}$ (circles; note the factor
of 4 in the definition) as calculated by DMFT for the one-band
Hubbard model with DOS (\ref{glferro1}) at $U=4$ and $n=0.58$.
Error-bars are of
the size of the symbols or smaller. The value of $\protect\chi %
_{F}^{-1}$ at $T=0.05$ is a data point, not an extrapolation. The
dotted line is a linear fit to $\protect\chi _{F}^{-1}$, the
dashed line a fit with a Brillouin function to $m$
\protect\cite{Ulmke98}. } \label{fig:figferro1}
\end{figure}

Investigations of the stability of metallic ferromagnetism on
fcc-type lattices within DMFT, in combination with
finite-temperature QMC techniques to solve the DMFT equations,
were first performed by Ulmke \cite{Ulmke98}. To detect a
ferromagnetic instability one calculates the temperature
dependence of the uniform static susceptibility $\chi _{F}$ from
the corresponding
two-particle correlation function \cite{Ulmke95a}. For $N_{%
\mbox{\scriptsize\em gfcc}}^{0}(E)$ and at an intermediate
interaction strength of $U=4$ the ferromagnetic response is found
to be strongest around quarter filling ($n\simeq 0.5$). The
susceptibility $\chi _{F}$ is seen to obey a Curie-Weiss law
(Fig.~\ref{fig:figferro1}).
Thus the Curie temperature $T_{c}$ can be safely extrapolated from the zero of $%
\chi _{F}^{-1}$. For $n=0.58$ this leads to a value of
$T_{c}=0.051(2)$. Below $T_{c}$ the magnetization $m$ (a
one-particle quantity determined by the local propagator) grows
rapidly, reaching more than 80\% of the fully polarized value
($m_{max}=n=0.58$) at the lowest temperature (30\% below $T_{c}$).
The three data points $m(T)$ in Fig.~\ref{fig:figferro1} are
consistent with a Brillouin function with the same critical temperature of $%
T_{c}=0.05$ and an extrapolated full polarization at $T=0$. (We
note that a saturated ground-state magnetization is also consistent
with the single-spin-flip analysis of the fully polarized state by
Uhrig \cite{Uhrig96}). So one finds {\em simultaneously} a
Curie-Weiss-type static susceptibility with
Brillouin-function-type magnetization, {\em and} a non-integer
magneton number, in qualitative agreement with experiments on 3$d$
transition metals. In the past, these features were attributed to
seemingly opposite physical pictures: the former to localized
spins, and the latter to itinerant electrons. However, this was
only because of the use of oversimplified, {\em static
}mean-field-type theories like Weiss mean-field theory (for spin
models) and Hartree-Fock (for electrons). Here we see that these
properties are natural features of {\em correlated} electronic
systems, which are generated by the quantum dynamics of the
many-body problem. Within DMFT this ``paradoxical'' behavior of
band ferromagnets is resolved without difficulty.

Collecting the values of $T_{c}(U,n)$ obtained by $\chi
_{F}^{-1}(T_c,U,n)=0$,
the boundaries between the ferromagnetic and the
paramagnetic phases are determined. Thereby one can construct the
$T$ vs. $n$ phase diagram for
different values of $U$. The region of stability increases with $U$~\cite%
{Ulmke98}.

\begin{figure}[tbp]
    \leavevmode
    \includegraphics[width=0.47\hsize]{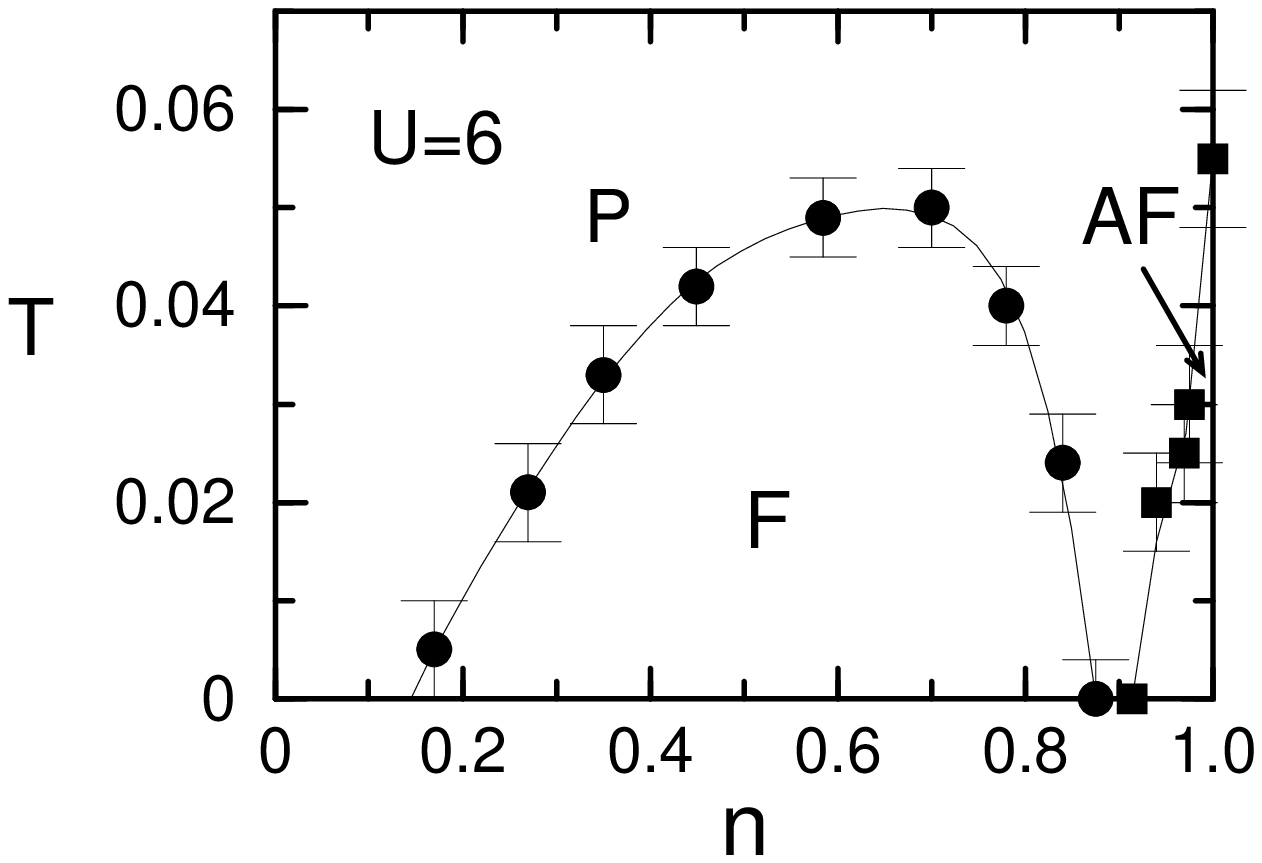}
    \hspace{0.05\hsize}
    \includegraphics[width=0.47\hsize]{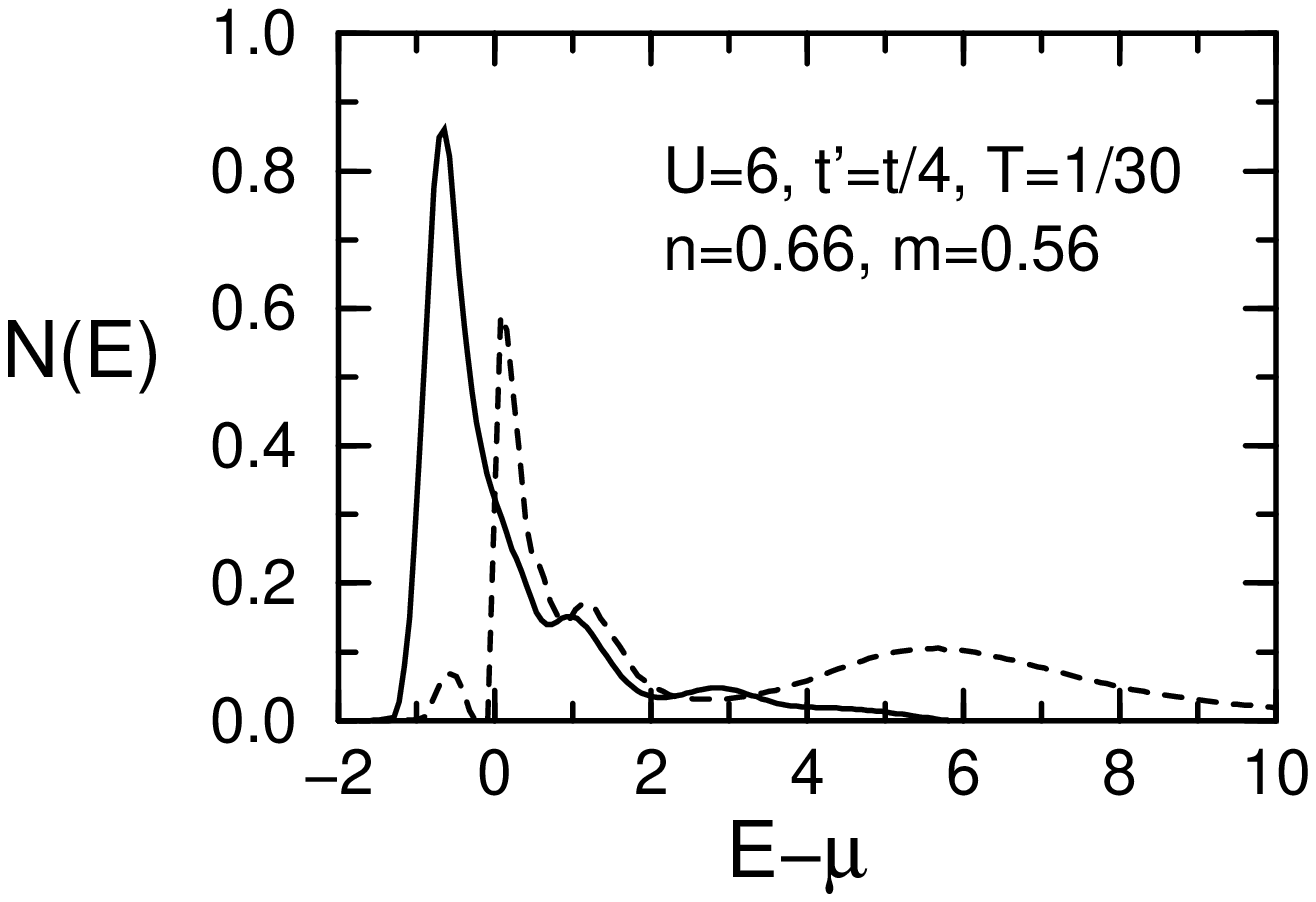}
    \par\vspace{-1.5ex}\makebox[0.6\hsize]
    {\small (a)}\makebox[0.4\hsize]{\small (b)}
\caption{(a) $T$ vs.~$n$ phase diagram of the Hubbard model as
obtained within DMFT for the DOS corresponding to a
three-dimensional fcc lattice with NN~hopping $t^{\prime}=t/4$
(see Fig.~\ref{fig:fccdos}); (b) DOS of the interacting electrons
in the ferromagnetic phase of (a), solid line: majority spin,
dashed line: minority spin \cite {Ulmke98}.} \label{fig:fcc3dim}
\end{figure}

To make contact with $d=3$ one can use the corresponding fcc DOS
shown in Fig.~\ref{fig:fccdos}. For $t^{\prime }=0$ no instability
is found at
temperatures accessible to QMC. However, already a small contribution of $%
t^{\prime }$-hopping (which is present in any real system) is
enough to stabilize a large region of metallic ferromagnetism in
addition to an antiferromagnetic phase (which is absent in
$d=\infty $), close to half filling (Fig.~\ref{fig:fcc3dim}a)~
\cite{Ulmke98}.
This shows the strong and subtle dependence of the stability on
the dispersion and the distribution of spectral weight in the DOS.
By developing increasingly refined schemes of iterated
perturbation theory to solve the DMFT equations Nolting and
collaborator~\cite{HerrVeg} obtained $T$ vs. $n$ phase diagrams
which reproduce the QMC results remarkably well.

We note that the maximal transition temperature is $T_{c}^{%
\mbox{\scriptsize max}}=0.05\simeq 500$~K for a band width $W$ = 4
eV. This is well within the range of real transition temperatures,
e.g., in nickel.

So far we only argued on the basis of the shape of the DOS of the
{\it non}-in\-ter\-act\-ing electrons, $N^{0}(E)$. On the other hand
the interaction will renormalize the band and relocate spectral
weight. Therefore it is not {\it a priori} clear at all whether
the arguments concerning the kinetic energy discussed in
subsection \ref{subs:routes} still hold at finite $U$. We
note that this effect is also not taken into account in any Stoner
theory. To settle this point one may calculate the DOS of the
interacting system, $N(E)$, by employing the maximum entropy
method for analytic continuation. In Fig.~\ref{fig:fcc3dim}b we
show $N(E)$ obtained with $N_{\mbox{\scriptsize\em gfcc}}^{0}(E)$
and the parameter values used to calculate
Fig.~\ref{fig:figferro1}. Clearly the ferromagnetic system is
metallic since there is appreciable weight at the Fermi level
($E=\mu $). Furthermore, the spectrum of the majority spins is
seen to be affected only slightly by the interaction, the overall
shape of the non-interacting DOS
being almost unchanged (the magnetization is quite large ($m=0.56$ at $%
n=0.66 $) and hence the electrons in the majority band are almost
non-interacting). This implies that the arguments concerning the
distribution of spectral weight in the non-interacting case and
the corresponding kinetic energy are even applicable to the
polarized, interacting case. The spectrum of the minority spins is
slightly shifted to higher energies and has a pronounced peak
around $E-\mu \simeq U=6$.

{}From Fig.~\ref{fig:fcc3dim}b we find the exchange splitting
$\Delta $ between the majority and minority bands to be $\Delta
=0.8$. Comparing the quantity $Um$ ($=3.4$ in the present case)
with $\Delta $ and $T_{c}$ we find a characteristic hierarchy of
energy scales:
\begin{equation}
Um>\Delta \gg T_{c}.  \label{hier}
\end{equation}
This is very different from results obtained by Stoner theory or
effective one-particle theories like LDA where all three
quantities are essentially equal. The generation of {\em small}
energy scales is a genuine correlation effect.

\begin{figure}[tbp]
    \includegraphics[width=0.56\hsize]{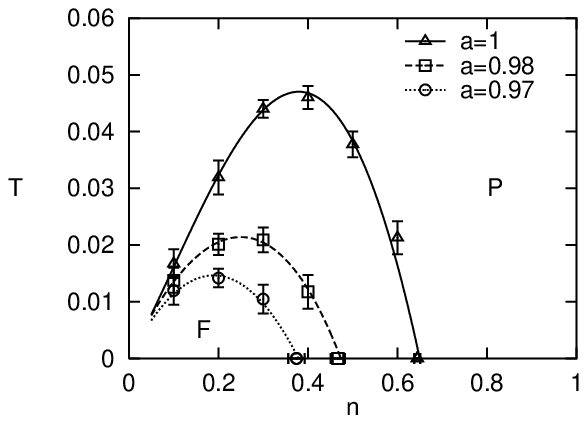}
    \includegraphics[width=0.45\hsize]{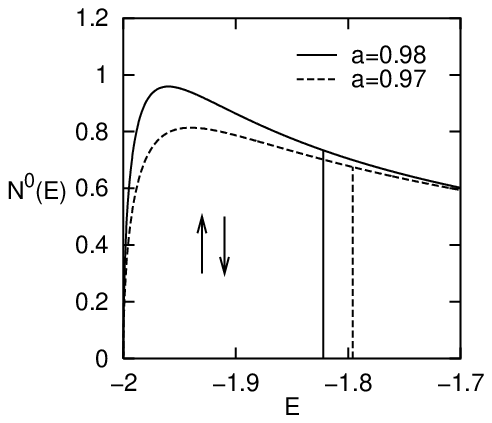}
    \par\vspace{-1.5ex}\makebox[0.6\hsize]
    {\small (a)}\makebox[0.4\hsize]{\small (b)}
 \caption{(a) $T$ vs.~$n$ phase diagram of the Hubbard model as obtained
within DMFT; (b) corresponding shapes of the non-interacting DOS;
Fermi energies for $n=0.3$ are indicated by vertical
lines~\protect\cite{Wahle98}.} \label{fig:adosdmft}
\end{figure}

To study the influence of the distribution of spectral weight on the
stability of ferromagnetism within the DMFT systematically, Wahle et
al.~\cite{Wahle98} recently solved the DMFT equations with a tunable
model DOS,
\begin{equation}
N^{0}(E)=c\,\frac{\sqrt{D^{2}-E^{2}}}{D+aE},  \label{eqn:dos}
\end{equation}%
with $c=(1+\sqrt{1-a^{2}})/(\pi D)$ and half-band width $D\equiv
2$. Here $a$ is an asymmetry parameter which can be used to change
the DOS continuously from a symmetric, Bethe lattice DOS ($a=0$)
to a DOS with a square-root divergence at the lower band edge
($a=1$), corresponding to a fcc lattice with $t^{\prime }=t/4$ in
$d=3$ (Fig.~\ref{fig:fccdos}). The strong dependence of the
stability of metallic ferromagnetism on the distribution of
spectral weight is shown in Fig.~\ref{fig:adosdmft}a. Already a
minute increase in spectral weight near the band edge of the
non-interacting DOS,
obtained by changing $a$ from $0.97$ to $0.98$ (see Fig.~\ref{fig:adosdmft}%
b) is enough to almost double the stability region of the
ferromagnetic phase. Obermeier et al.~\cite{Obermeier97} found
ferromagnetism even on a hypercubic, i.e.,~bipartite, lattice, but
only at very large $U$ values ($U>30$).

\begin{figure}[tbp]
\begin{center}
\includegraphics[width=0.6\hsize]{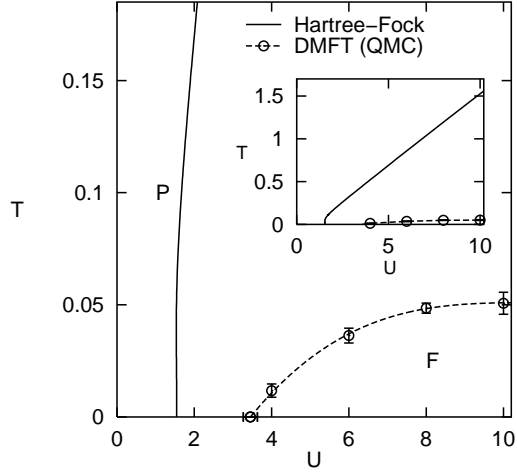}
\end{center}
\caption{$T$ vs.~$U$ phase diagram for a strongly peaked DOS
($a=0.98$, see Fig.~\ref{fig:adosdmft}b) at $n=0.4$ within DMFT
(circles; dashed line is
guide to the eyes only) in comparison with Hartree-Fock (solid line)~ %
\protect\cite{Wahle98}.} \label{fig:adostvsu}
\end{figure}

The importance of genuine correlations for the stability of
ferromagnetism is apparent from Fig.~\ref{fig:adostvsu}, where the
DMFT results are compared with Hartree-Fock theory~\cite{Wahle98}.
The quantum fluctuations, absent in Hartree-Fock theory, are seen
to reduce the stability regime of ferromagnetism drastically.
Spatial fluctuations (e.g., spin waves), absent in the DMFT, may
reduce the stability regime further.
\subsection{Additional Interactions}

\label{sec:nninteractions}

In the one-band Hubbard model only the local interaction is
retained. Thereby several interactions which naturally arise when
the Coulomb interaction is expressed in Wannier representation are
neglected. Even in the limit of a single band and taking into
account only NN~contributions,
four additional interactions appear~\cite%
{Hubbard63,GammCamp,Hirsch,Vollhardt97b+99}:
\begin{eqnarray}
{V}_{\mbox{\scriptsize 1-band}}^{\mbox{\scriptsize NN}} &=&\sum_{%
\mbox{\scriptsize NN}}\bigg[V{n}_{i}{n}_{j}+X\sum_{\sigma
}({c}_{i\sigma
}^{\dagger }{c}_{j\sigma }^{\phantom\dagger }+{\rm h.c.})({n}_{i,-\sigma }+{n%
}_{j,-\sigma })  \nonumber \\
&&-2F({\bf S}_{i}\cdot {\bf S}_{j}+\frac{1}{4}{n}_{i}{n}_{j})+F{^{\prime }}({%
c}_{i\uparrow }^{\dagger }{c}_{i\downarrow }^{\dagger }{c}_{j\downarrow }^{%
\phantom\dagger }{c}_{j\uparrow }^{\phantom\dagger }+{\rm
h.c.})\bigg]. \label{vnn}
\end{eqnarray}%
Here the first term corresponds to a density-density interaction,
the second term to a density-dependent hopping, and the fourth
term describes the hopping of doubly occupied sites. In
particular, the third term (with $F$ $=$ $F^{\ast }/Z>0$)
\begin{equation}
H_{F}=-2\frac{F^{\ast }}{Z}\sum_{\mbox{\scriptsize NN}}{\bf
S}_{i}\cdot {\bf S}_{j}
\end{equation}%
describes the direct ferromagnetic exchange between electrons on
NN~sites. It is this interaction which Heisenberg singled out in
his original spin-model as the main source of ferromagnetism. It
should be noted, however, that this interaction is present even
when the electrons are free to move. The exchange interaction will
be quite small, but nevertheless it favors ferromagnetic ordering
in the most obvious way. Hirsch~\cite{Hirsch} argued that this
term is the main driving force for metallic ferromagnetism in
systems like iron, cobalt, and nickel. Indeed, one can show
rigorously that a next-neighbor direct exchange interaction, if
only chosen large enough, easily triggers the ferromagnetic
instability~\cite{StraKoll,Vollhardt97b+99,Wahle98}.
Hence the NN~exchange may well be important for systems on the
verge of a ferromagnetic instability.

\begin{figure}[tbp]
\begin{center}
\includegraphics[width=0.6\hsize]{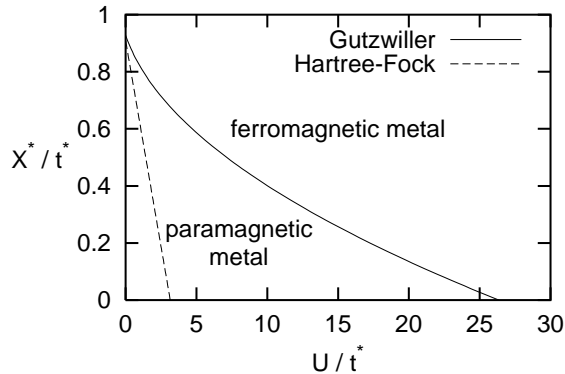}
\end{center}
\par
\caption{Bond-charge interaction $X$ vs.~$U$ phase diagram for the
generalized model $H=H_{\mbox{\scriptsize Hub}}+H_{X}$ within the
Gutzwiller
approximation for the semi-elliptical Bethe DOS at $n=0.9$~\protect\cite%
{kollar00}.} \label{fig:xterm}
\end{figure}

Another term that can be of importance for the stabilization of
ferromagnetism is the bond-charge interaction ($X$ $=$ $X^{\ast
}/\sqrt{Z}$)
\begin{equation}
H_{X}=\frac{X^{\ast }}{\sqrt{Z}}\sum_{\mbox{\scriptsize
NN}\sigma }({c}_{i\sigma }^{\dagger }{c}_{j\sigma }^{\phantom\dagger }+{\rm %
h.c.})({n}_{i,-\sigma }+{n}_{j,-\sigma }).
\end{equation}%
It effectively gives rise to correlated hopping, i.e., the hopping
amplitude of an electron now depends on the presence of electrons
of the opposite spin polarization. The magnitude of $X$ has been
estimated to be of order 0.1-1~eV~\cite{Hubbard63,GammCamp}, and
hence $X$ may be comparable to the
hopping $t$, although typically smaller than $U$. Hartree-Fock treatment of $%
H$ $=$ $H_{\mbox{\scriptsize Hub}}$ $+$ $H_{X}$ shows that
correlated hopping can lead to a
spin-dependent band narrowing which may stabilize ferromagnetism~\cite%
{amadon96}. We note that in DMFT $H_{X}$ does not reduce to its
Hartree-Fock contribution. The correlation effects introduced by
$H_{X}$ were recently studied within the Gutzwiller approximation
(GA)~\cite{kollar00}. The GA yields the exact evaluation of
expectation values in terms of the Gutzwiller wave function in the
limit of $d$ $\rightarrow $ $\infty $~\cite{Metzner87,Metzner89a}
and goes beyond Hartree-Fock theory by including
correlations explicitly. It was found that correlated hopping with $X$ $>$ $%
0 $ can lower the critical value of $U$ for ferromagnetism
considerably as shown in Fig.~\ref{fig:xterm}.
Compared to Hartree-Fock theory the ferromagnetic region of the
phase diagram is seen to be much reduced. Recently it was also
shown how to incorporate $H_{X}$ into DMFT~\cite{schiller99}, at
least in principle. It will be interesting to see how the GA
results compare with future DMFT calculations for $H$ $=$
$H_{\mbox{\scriptsize Hub}}$ $+$ $H_{X}$.

Clearly, NN~interactions may help to stabilize ferromagnetism.
However, since the considerably larger Hubbard interaction $U$,
together with a suitable kinetic energy, is already sufficient to
trigger a ferromagnetic instability, the ferromagnetic exchange
and bond-charge interactions appear to play only a secondary role.


\section{Band Degeneracy and Local Exchange}

\label{sec:zweiband}

\begin{figure}[tbp]
\begin{minipage}{.3\textwidth}
\includegraphics[width=\textwidth]{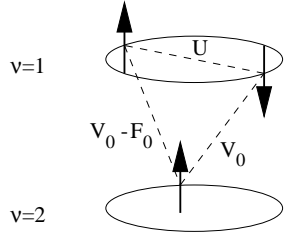}
\end{minipage}\hfill
\begin{minipage}{.6\textwidth}
\caption{Illustration of the local interactions (Hubbard $U$ and
Hund's rule couplings $V_0$, $F_0$ defined in
(\ref{eqn:zweiband})) between electrons in a two-band
model.\label{fig:hundillu}}
\end{minipage}
\end{figure}

\begin{figure}[tbp]
 \vspace{.8cm}

\setlength{\unitlength}{0.00244\textwidth}
\begin{picture}(410,62)
     \linethickness{1pt}
     {
       \put(60,37){\line(1,0){20}}
       \put(90,37){\line(1,0){20}}
       \put(160,37){\line(1,0){20}}
       \put(190,37){\line(1,0){20}}
       \put(260,37){\line(1,0){20}}
       \put(290,37){\line(1,0){20}}
       \put(360,37){\line(1,0){20}}
       \put(390,37){\line(1,0){20}}
       \put(60,57){\line(1,0){20}}
       \put(90,57){\line(1,0){20}}
       \put(160,57){\line(1,0){20}}
       \put(190,57){\line(1,0){20}}
       \put(260,57){\line(1,0){20}}
       \put(290,57){\line(1,0){20}}
       \put(360,57){\line(1,0){20}}
       \put(390,57){\line(1,0){20}}
       \put(0,54){$\nu=1$}
       \put(0,32){$\nu=2$}
       }
     {
       \put(66,54){\Large $\uparrow$}
       \put(96,54){\Large $\uparrow$}
       \put(166,54){\Large $\uparrow$}
       \put(196,50){\Large $\downarrow$}
       \put(266,34){\Large $\uparrow$}
       \put(296,50){\Large $\downarrow$}
       \put(366,34){\Large $\uparrow$}
       \put(396,54){\Large $\uparrow$}
       }
     {
       \put(0,5){$\Delta E\! =\! $}
       \put(72,5){$ 0$}
       \put(166,5){$ -2t^2/U$}
       \put(260,5){$ -2t^2/V_0$}
       \put(345,5){{$-2t^2/(\!V_0\!-\!F_0$)}}
       }
   \end{picture}
\caption{Energy gain in second-order perturbation theory in $t$
for four configurations with two electrons on two sites. Most
favorable is ferromagnetism combined with an alternating
occupation of orbitals. Note,
that the true ground state is not the last configuration above, i.e., $%
c^\dagger_{ i=1 \protect\nu=1 \uparrow} c^\dagger_{2 2 \uparrow}
|0\rangle$, but the corresponding orbital singlet $\frac{1}{2} (
c^\dagger_{ 1 1 \uparrow} c^\dagger_{2 2 \uparrow} - c^\dagger_{1
2 \uparrow} c^\dagger_{2 1 \uparrow} ) |0\rangle$. }
\label{fig:superexchange}
\end{figure}
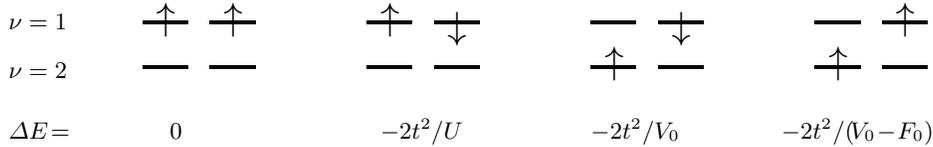

Besides the NN~Heisenberg exchange interaction another much larger
exchange term is present in ferromagnets like iron, cobalt, and
nickel, namely the local exchange between electrons in {\em
different} orbitals on the same lattice site. It has long been
speculated that this exchange interaction, which is known to align
electrons on isolated atoms (Hund's first rule), may also lead
to ferromagnetism in the bulk, being transmitted by the kinetic energy \cite%
{Slater36}. The simplest model for this mechanism is the two-band
Hubbard model with local exchange $F_{0}$ and Coulomb repulsion
$V_{0}$ between two orbitals $\nu =1,2$ (Fig.~\ref{fig:hundillu}):
\begin{eqnarray}
H_{\mbox{2-band}} &=&-t\sum_{\mbox{\scriptsize NN,}\sigma \nu
}{c}_{i\nu
\sigma }^{\dagger }{c}_{j\nu \sigma }^{\phantom{\dagger}}+U\sum_{i\nu }{n}%
_{i\nu \uparrow }{n}_{i\nu \downarrow }  \nonumber \\
&&+\sum_{i\sigma \sigma ^{\prime }}(V_{0}-\delta _{\sigma \sigma
^{\prime }}F_{0}){n}_{i1\sigma }{n}_{i2^{\prime }\sigma ^{\prime
}}-F_{0}\!\sum_{\sigma \neq \sigma ^{\prime }}\!{c}_{i1\sigma }^{\dagger }{c}%
_{i1\sigma ^{\prime }}^{\phantom{\dagger}}{c}_{i2\sigma ^{\prime
}}^{\dagger }{c}_{i2\sigma }^{\phantom{\dagger}}.
\label{eqn:zweiband}
\end{eqnarray}
This two-band model \cite{2Band1} and its multi-band generalizations \cite%
{2Band2} were studied in considerable detail by various
theoretical techniques \cite{NoltingPAM}. Recent investigations
\cite{2Band3} were triggered by the renewed interest in the
electronic properties of transition metal oxides where the doubly
degenerate $e_{g}$ bands of the $d$~electrons plays a very
important role. At quarter filling (one electron per site; $n=1$),
ferromagnetism can be understood by superexchange within
strong-coupling perturbation theory
(Fig.~\ref{fig:superexchange}). But away from quarter (or half)
filling, the model is much more difficult to treat. In this regime
DMFT, solved by QMC, once more provides a powerful tool to
investigate this model \cite{Held98}, at least, if the last term
in (\ref{eqn:zweiband}), i.e., the spin-flip contribution of
$F_{0}$, is neglected.

The calculated $T$ vs. $n$ phase diagram is shown in Fig.~\ref%
{fig:zweibandphasdiag} for a symmetric Bethe DOS. Without Hund's
rule coupling $F_{0}$ ferromagnetism is {\em unstable} for this
DOS, at least at moderate values of $U$ (see
Sec.~\ref{sec:asymmetricDOS} for the one-band model and
\cite{Held98} for the two-band model). However, when the Hund's
rule exchange is included a metallic ferromagnetic phase is
stabilized between an alternating (staggered) orbital ordered
state at quarter filling and antiferromagnetism at half filling.

While orbital ordering at $n=1$ and antiferromagnetism at $n=2$
can be understood by superexchange, the mechanism for metallic
ferromagnetism is more subtle. Of course, the virtual
superexchange processes described above are still present. But the
additional electrons doped to the $n=1$ system may now move freely
between two singly occupied sites, i.e., are unhindered by the
Coulomb repulsion $U$ or $V_{0}.$ Furthermore, if the electrons on
the two
sites are spin aligned they do not even have to spend the exchange energy $%
F_{0}$. Therefore a ferromagnetic alignment of the spins improves
the kinetic energy of the electrons. This is essentially the
concept of {\em double exchange} introduced by Zener
\cite{Zener51b} to explain ferromagnetism in manganites such as
La$_{1-x}$Ca$_{x}$MnO$_{3}$, and put on a firmer theoretical
fundament by Anderson and Hasegawa \cite{Anderson55a}.

\begin{figure}[tbp]
\begin{center}
\includegraphics[width=0.6\textwidth]{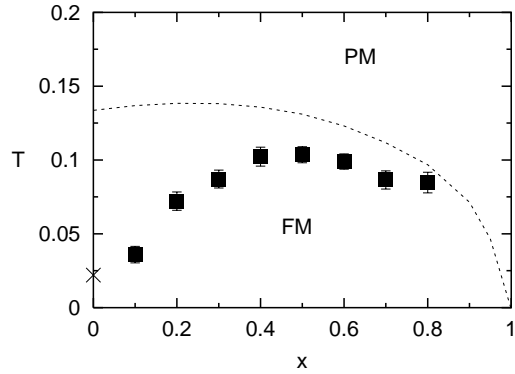}
\end{center}
\caption{Curie temperature $T_c$ for the phase transition from the
paramagnetic (PM) to the ferromagnetic (FM) phase as a function of
Ca doping $x$ ($n=1-x$). Dashed line: KLM with Bethe DOS (band
width $W=2$) and Hund's rule coupling $J=3/2$; squares: correlated
electron model (\ref{eqn:manganites}) which also takes into
account the Coulomb interaction between $e_{g}$ electrons ($U=8$,
$V_0=6$, and $F_0=1$); cross: Weiss mean-field theory
(superexchange) for this model~\protect\cite{Held00a}. Note that,
without
the coupling to the $t_{2g}$ spin, no ferromagnetism is observed in Fig.~\ref%
{fig:zweibandphasdiag} for $n<1$ at similar interaction strengths.
} \label{fig:manganitephasdiag}
\end{figure}

In manganites, the cubic crystal field splits the five Mn $d$
orbitals into three $t_{2g}$ and two $e_{g}$ orbitals. The former
have a lower energy and hybridize less strongly with the O
$p$ orbitals. Thus, the three electrons within the $t_{2g}$
orbitals can be approximately described by a localized spin ${\bf
S}_{i}$, with the remaining \mbox{$1\!-\!x$} electrons occupying
the $e_{g}$ orbitals. If the Coulomb repulsion between $e_{g}$
electrons is taken into account one arrives at a correlated
electron model for manganites:
\begin{equation}
H=H_{\mbox{\scriptsize 2-band}}-2J\sum_{\nu =1}^{2}\sum_{i}{\bf s}_{i\nu }\cdot {\bf S}%
_{i\nu }.  \label{eqn:manganites}
\end{equation}%
Here, $H_{\mbox{\scriptsize 2-band}}$ is defined in
(\ref{eqn:zweiband}),
${\bf s}_{i\nu }$=$\frac{1}{2}\sum_{\sigma \sigma ^{\prime
}}{c}_{i\nu \sigma }^{\dagger }{\mathbf \tau }_{\sigma \sigma
^{\prime }}{c}_{i\nu \sigma ^{\prime }}^{\phantom{\dagger}}$
denotes the $e_{g}$ spin
(${\mathbf \tau}$: Pauli matrices), and $J$ is the local exchange between $%
t_{2g}$ and $e_{g}$ electrons. Without Coulomb interaction ($%
U=V_{0}=F_{0}=0)$, the Hamiltonian (\ref{eqn:manganites}) reduces
to the ferromagnetic Kondo lattice model (KLM) which was
investigated intensively in recent years \cite{KLM}. This model
forms the theoretical basis for the double exchange mechanism: at
$J\gg t$, the optimization of the kinetic
energy of the $e_{g}$ electrons requires a ferromagnetic environment of $t_{{%
2g}}$ spins. The KLM fails to describe the correct behavior for $x
\lesssim 0.5$ since it does not penalize double occupations, i.e.,
two $e_{g}$ electrons on the same site. The suppression of double
occupations induced by the Coulomb repulsions $U$ and $V_{0}$ was
investigated
within DMFT in \cite%
{Held00a}, and was shown to lead to a {\em crossover} from double
exchange to superexchange. This results in a maximum in the Curie
temperature (Fig.~\ref{fig:manganitephasdiag}) in qualitative
agreement with experiment \cite{Schiffer95a}.
\begin{figure}[tbp]
\begin{center}
\includegraphics[width=0.6\textwidth]{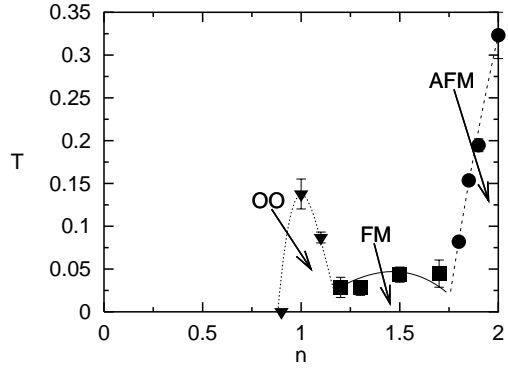}
\end{center}
\caption{$T$ vs.~$n$ phase diagram of the two-band Hubbard model
including antiferromagnetism (AF), ferromagnetism (FM), and
orbital ordering (OO) for
a Bethe DOS (total width $W=4$), $U=6$, $V_0=4$, and $F_0=2$ ~\protect\cite%
{Held98}.} \label{fig:zweibandphasdiag}
\end{figure}

\section{Conclusion and Outlook}

\label{sec:concl} In this paper we reviewed recent developments in
our understanding of the origin of metallic ferromagnetism in the
one-band Hubbard model and in band-degenerate models on the basis
of the dynamical mean-field theory (DMFT). In the one-band Hubbard
model metallic ferromagnetism is found to exist in a surprisingly
large region of the $U$ vs. $n$ phase diagram. A stabilization of
this phase at intermediate $U$ values requires a sufficiently
large spectral weight near one of the band edges
as is typical for frustrated (e.g., fcc-type) lattices which
optimize the kinetic energy of the polarized state and, at the
same time, frustrate the parasitic antiferromagnetic order. This
finding, together with the results obtained for dimension $d=1$
\cite{MuellerHartmann95,Pieri96,Penc96,Daul}, finally establishes
the stability of band ferromagnetism in the one-band Hubbard model
on regular lattices and at intermediate values of the interaction
$U$ and density $n$. Thereby one of the prominent questions of
many-body theory in this field is finally answered.

By contrast, the origin of metallic ferromagnetism in the
band-degenerate Hubbard model at intermediate $U$ values is not
primarily a spectral weight effect but is already caused by
moderately strong Hund's rule couplings . In this respect the
origin of ferromagnetism in the orbitally degenerate model is more
straightforward than that in the (less realistic) one-band case.
Nevertheless, in the absence of orbital ordering the resulting
magnetic phase diagrams are remarkably similar. The identification
of a single main driving force for the stabilization of metallic
ferromagnetism in the one-band and the band-degenerate Hubbard
model, respectively, helps to differentiate between different
mechanisms. However, in real systems these effects will tend to
conspire, as is evident, for example, in nickel where an fcc
lattice leads to a strongly asymmetric DOS and the band degeneracy
brings with it Hund's rule couplings.

These insights also helped to understand the origin of itinerant
ferromagnetism in more complicated models, e.g., the ferromagnetic
Kondo lattice model with and without Coulomb correlations, which
is employed to
understand the physics of manganites with perovskite structure, like La$%
_{1-x}$(Sr,Ca)$_{x}$MnO$_{3}$. It was found that in this model
double exchange can explain ferromagnetism only for doping $x
\gtrsim 0.5$. At lower values of $x$ the suppression of double
occupations by the local Coulomb repulsion becomes more and more
important and leads to a crossover from double exchange to
superexchange. This results in a maximum of the Curie temperature
in qualitative agreement with experiment.

As discussed in the Introduction the problem of metallic
ferromagnetism was, until recently, investigated by two
essentially separate communities -- one using model Hamiltonians
together with many-body techniques, the other employing density
functional theory (DFT) in the local density approximation (LDA).
In view of the individual power of LDA and the model Hamiltonian
approach it is highly desirable to combine these techniques,
thereby creating an enormous potential for all future {\em ab
initio} investigations of real materials, including, e.g.,
$f$-electron systems, Mott insulators and metallic ferromagnets. A
combination of these two approaches had already been used to
investigate band ferromagnetism some time ago \cite{Nolting}.
Recently, a fusion has started to emerge in new directions. One is
the construction of multi-band Gutzwiller wave functions in
combination with spin-density functional theory in the limit of
large coordination numbers \cite{Weber}. It was already
successfully applied to ferromagnetic nickel, leading to
significant improvements over LDA results. The other is the {\em
ab initio} computational scheme LDA+DMFT \cite{Held00} which was
recently used to investigate transition metal oxides and to
calculate the magnetic excitation spectrum of ferromagnetic iron
\cite{Lichtenstein}. Without doubt these and related methods
\cite{NoltingAuger} will rapidly develop into standard tools for
future investigations of band ferromagnetism and other electronic
correlation phenomena.

\section*{Acknowledgments}

We thank M. Ulmke and J. Wahle for fruitful discussions. This work
was supported in part by the Sonderforschungsbereich 484 of the
Deutsche Forschungsgemeinschaft and by a Feodor-Lynen grant of the
Alexander von Humboldt-Foundation (KH).

\end{document}